# Quantifying myelin in brain tissue using color spatial light interference microscopy (cSLIM)

Michael Fanous[†,‡], Megan P. Caputo[○], Young Jae Lee[†,§], Laurie A. Rund[∥], Catherine Best-Popescu[§], Mikhail E. Kandel[†,⊥], Rodney W. Johnson[○,∥], Tapas Das[#], Matthew J. Kuchan[∇] & Gabriel Popescu[†,‡,⊥]

[†]Quantitative Light Imaging Laboratory, Beckman Institute for Advanced Science and Technology, University of Illinois at Urbana-Champaign, Urbana, Illinois 61801, USA

[‡] Department of Bioengineering, University of Illinois at Urbana-Champaign, 306 N. Wright Street, Urbana, IL 61801, USA

[○]Division of Nutritional Sciences, University of Illinois, Urbana-Champaign, Illinois 61801, USA

[§]Neuroscience Program, University of Illinois at Urbana-Champaign, Urbana, Illinois 61801, USA

[∥] Laboratory of Integrative Immunology & Behavior, Department of Animal Sciences, University of Illinois, Urbana-Champaign, Illinois 61801, USA

[⊥]Department of Electrical and Computer Engineering, University of Illinois at Urbana-Champaign, 306 N. Wright Street, Urbana, IL 61801, USA

[#]Abbott Nutrition, Discovery Research, Columbus, OH 43219, USA

[∇]Abbott Nutrition, Strategic Research, 3300 Stelzer Road, Columbus, OH, USA

**ABSTRACT:** Deficient myelination of the brain is associated with neurodevelopmental delays, particularly in high-risk infants, such as those born small in relation to their gestational age (SGA). New methods are needed to further study this condition. Here, we employ Color Spatial Light Interference Microscopy (cSLIM), which uses a brightfield objective and RGB camera to generate pathlength-maps with nanoscale sensitivity in conjunction with a regular brightfield image. Using tissue sections stained with Luxol Fast Blue, the myelin structures were segmented from a brightfield image. Using a binary mask, those portions were quantitatively analyzed in the corresponding phase maps. We first used the CLARITY method to remove tissue lipids and validate the sensitivity of cSLIM to lipid content. We then applied cSLIM to brain histology slices. These specimens are from a previous MRI study, which demonstrated that appropriate for gestational age (AGA) piglets have increased internal capsule myelination (ICM) compared to small for gestational age (SGA) piglets and that a hydrolyzed fat diet improved ICM in both. The identity of samples was blinded until after statistical analyses.

Myelin is a critical component of the nervous system white matter. The myelin sheath surrounds axons and thereby provides the necessary insulation for the efficient transmission of neural electrical signals across brain regions[1]. Myelination of fiber bundles is one of the slowest processes of brain maturation in humans, starting at 16 weeks gestational age and increasing rapidly from 24 weeks through to the perinatal period and continuing until puberty[2]. The rapid growth phase during the perinatal period represents a vulnerable time for neural development and is a critical period for small for gestational age (SGA) infants. Myelin formation plays a crucial role in the operation of diverse cognitive faculties during the perinatal period[3-7] and continuing throughout adulthood[8].

Particularly impacted by deficient myelination are intrauterine growth-restricted (IUGR) and low birth weight (LBW) infants. These infants are at higher risk of morbidity and mortality early in life[9] and display adverse neurological outcomes that include developmental disorders, learning and attention difficulties, behavioral issues, neuropsychiatric disorders and epilepsy that persist into adulthood[10-16]. Approximately 24% of newborn human infants are IUGR, and more than 20 million infants are born each year with LBW[17,18]. Currently, there are no available therapeutic interventions to prevent or treat brain damage in the IUGR newborn. Early dietary interventions aimed at mitigating the cognitive deficits associated with IUGR and LBW are important given that rapid growth occurs early in postnatal life.

Multiple modalities have been used to assess myelin deposition in biological samples. Luxol Fast Blue (LFB) is a dye that stains myelin in formalin fixed tissue[19]. LFB provides information on the presence of myelin in terms of its spatial distribution but does not allow quantification. Magnetic resonance imaging (MRI) at high field strength allows in vivo visualizations of human brain structures[20] and has simplified the analysis of myelin concentrations[21]. However, MRI detection is indirect, relying on the proton spins of water molecules. Although a reasonable agreement between MRI and LFB staining has been shown, MRI remains a low sensitivity method for myelin imaging[22].

Other efforts to quantify myelin in brain tissue have included the use of the proton-induced X-ray emission (PIXE), which generates X-rays with an energy spectrum that is characteristic of a chemical component[21]. This enables a quantitative determination of the spatial distribution of components within a sample. However, its resolution is relatively poor and it involves tedious calculations and cumbersome equipment[20].

Here, we use our recently devised color spatial light interference microscopy (cSLIM)[23] as a quantitative phase imaging (QPI) technique[24-34] to quantify myelin in brain tissue. QPI is a technique that can evaluate nanometer scale pathlength changes in biological specimens. The data in the phase image, $\varphi(x, y)$, is a numerical assessment of the nanoarchitecture of the tissue biopsy.

Traditional interferometric methods use coherent light sources, which obstruct the contrast in images with speckle artifacts[35,36]. cSLIM overcomes this issue by exploiting a white light illumination, which averages the speckles and, overall, boosts the phase sensitivity[37-40]. The SLIM principle of operation relies on phase shifting interferometry applied to a phase contrast geometry. Thus, we shift the phase delay between the incident and scattered field in increments of π/2 and acquire 4 respective intensity images, which is sufficient to extract the phase image unambiguously.

Because cSLIM uses a brightfield objective and RGB camera, one of the 4 intensities is the standard LFB color image. Thus, simultaneously, cSLIM yields both a brightfield image and a phase map, as illustrated in Figure 1. The contrast in the cSLIM (phase) images are produced by the tissue refractive index spatial inhomogeneity, which is an intrinsic morphological marker with relevance to key disease features[41,42]. Unlike previous QPI methods applied for diagnosis, cSLIM is employed on already stained tissue slides, can scan an entire biopsy slide quickly, and generate the brightfield image and corresponding phase map. Furthermore, the effects of the stain on the phase image are normalized, using a procedure reported earlier[43]. One important application of cSLIM is to use the stain map from the brightfield image to create a binary mask, which is then applied to the phase image. This way, the intrinsic tissue architecture and density are analyzed on specific regions of interest, as revealed by the stain, yet, free of stain artifacts.

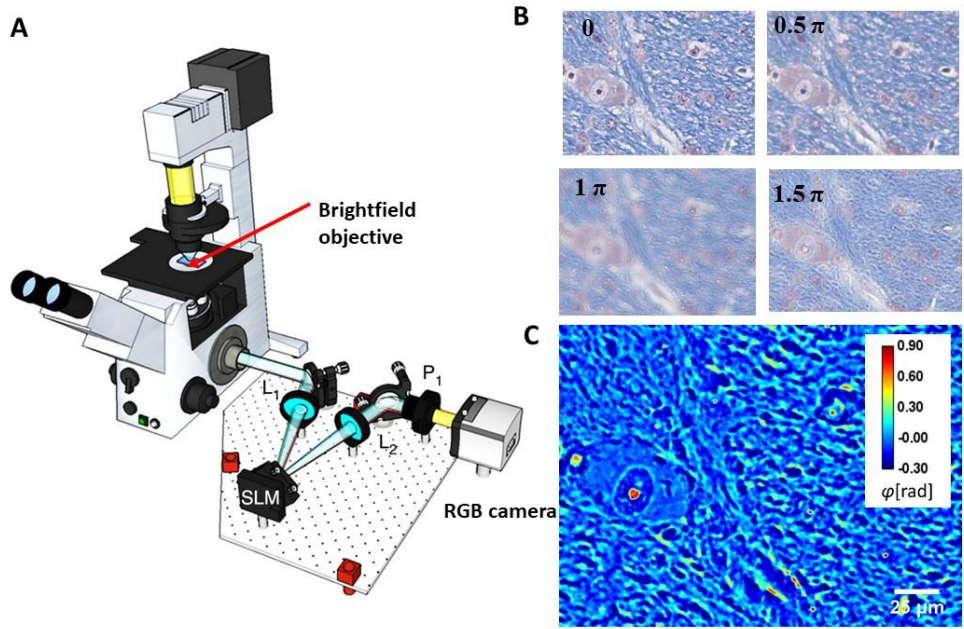

**Figure 1**. Schematic setup for cSLIM. (**A**) The cSLIM module is attached to a commercial phase contrast microscope (Axio Observer Z1, Zeiss), and uses a brightfield objective with an RGB camera. (**B**) The four phase-shifted color interferograms, with the initial unshifted frame corresponding to a brightfield image. (**C**) Computed SLIM image.

In an effort to first probe our technique's sensitivity, we employed the CLARITY method[44] on mouse brain tissue. Clarity removes the lipid component of myelin and renders the brain tissue transparent. The resultant cleared brain-hydrogel hybrid is physically stable and maintains the internal structure of proteins and nucleic acids isotropically. Comparing measurements before and after applying the clarity procedure allowed myelin lipids to be assessed specifically in terms of absolute dry mass values[44].

Next, we applied our cSLIM approach to analyze myelin content in piglet brain tissue. Piglets are useful to study neurodevelopmental disorders, because like humans, they are gyrencephalic and they share common brain growth patterns. In order to validate our method, we used two variables: gestational age and diet. Previously, a proprietary blend of dietary hydrolyzed fats (HF) increased relative volumes of 7 brain regions and white matter volume measured by MRI in both SGA and AGA piglets compared to a control fat blend[43]. We analyzed brain sections from these piglets using cSLIM while blind to the treatments. The results show that our method recapitulated the impact of both HF and gestational age on the HF myelin content in the internal capsule.

Tissues were obtained from piglets used in a previous study[45]. Piglets were PIC Camborough (dam) x PIC 359 (sire), full term, naturally delivered, sex-matched littermate pairs (n = 18 AGA, n = 18 SGA) obtained from the University of Illinois Swine Farm at 2 days old to allow colostrum consumption. Piglets underwent minimal routine processing on the farm (e.g., males were not castrated), but received iron dextran (Henry Schein Animal Health, Dublin, OH, USA) and antibiotic injection (Gentamicin Piglet Injection, Agri Laboratories, Ltd., St. Joseph, MO, USA) per routine farm practice and according to label. SGA was defined as piglets weighing 0.5-0.9 kg at birth, and AGA was defined as piglets weighing 1.2-1.8 kg at birth. Piglets were placed individually into a caging system under standard conditions as described in a previous publication[46] and randomly assigned to HF or CON diet treatment groups in a 2 x 2 factorial arrangement of size (AGA or SGA) and diet (CON or HF). All animal care and experimental procedures were in accordance with the National Research Council Guide for the Care and Use of Laboratory

Animals and approved by the University of Illinois at Urbana-Champaign Institutional Animal Care and Use Committee.

Diets were formulated to meet the nutritional needs of neonatal piglets[47] and were supplied in a premixed, ready to feed format by Abbott Nutrition (Columbus, OH, USA). CON formula contained 100% triglyceride rich oil as a fat source. HF formula contained a proprietary blend of soy free fatty acids, monoacuglycerol palmitate, and phospholipids. In HF diet, 50% triglyceride oil was replaced with HF blend. Piglets were weighed each morning and provided the liquid diet (300 mL formula/kg body weight/d) in 5 equal bolus feedings given at 09:00, 13:00, 16:00, 19:00, and 22:00. Supplemental water as not provided aside from that in the diet. At 26-29 days of age, piglets were anesthetized (telazol:ketamine:xylazine cocktail (100/50/50 mg/kg; 0.022 mL/kg body weight i.m.)) and then euthanized by intracardiac injection of sodium pentobarbital (Fatal Plus; 72 mg/kg body weight). Brains were quickly extracted, and the left hemisphere placed in 10% neutral buffered formalin (NBF; Leica Biosystems, Wetzlar, Germany) for fixation, refreshing the formalin solution after 24 hours to ensure adequate fixation.

The brains were cut in coronal sections 5 mm thick per block beginning 0.5 mm rostral to the optic chiasm. Blocks were soaked in 10% NBF overnight before processing. Fixed tissue blocks were then cut into 4 μm thick sections, mounted on glass slices and subsequently stained with LFB by the Veterinary Diagnostic Laboratory (University of Illinois Urbana-Champaign, Urbana, IL, USA) according to standard protocols for formalin-fixed, paraffin embedded brain and spinal cord tissue sections. Following LFB staining, slides were selected for imaging. Hyper-stained slides were omitted from imaging analysis as intense cresyl violet background staining may have interfered with cSLIM measurement. Final analysis included 10 AGA slides (5 HF fed and 4 CON fed), and 9 SGA (5 HF and 4 CON) slides. For all the imaging and analysis, we were blind to the sample diet categories. The internal capsule regions were selected for analysis because the area is myelin-rich myelin, as indicated by the dense dye regions (dark blue, in Figure 2A).

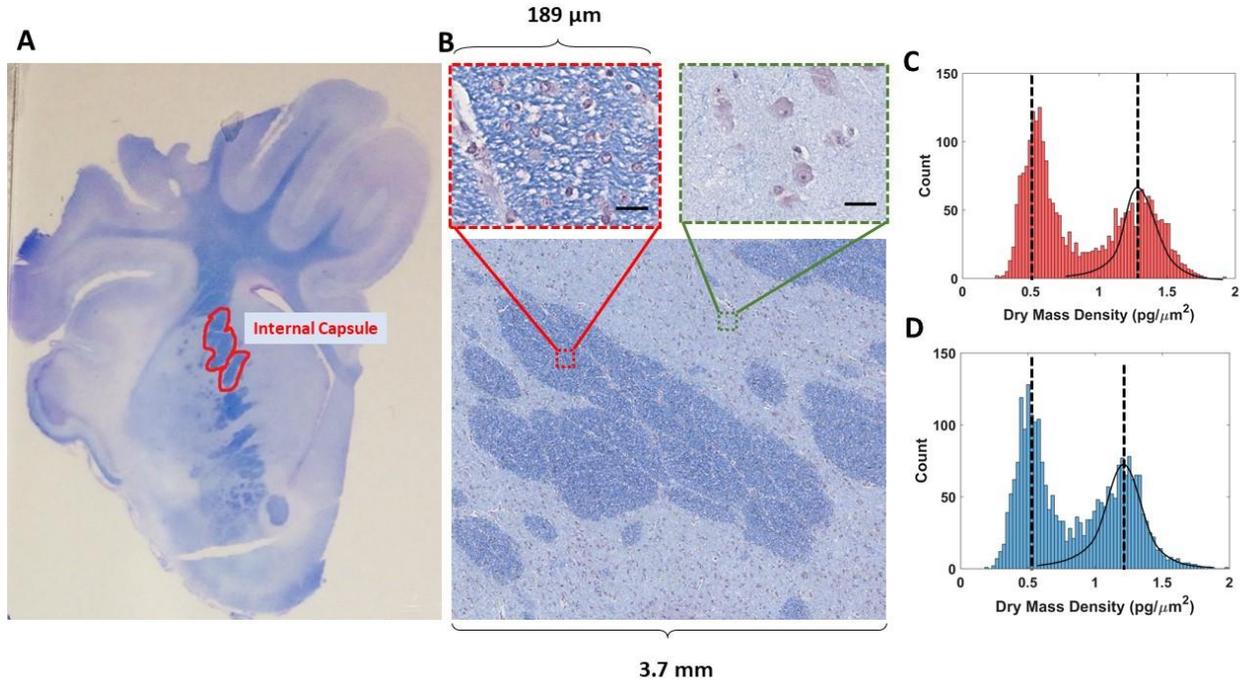

**Figure 2.** Picture of a slide of a piglet brain tissue with the internal capsule delineated in red (**A**). A stitch of brightfield cSLIM images with selected single frames of high and low myelin regions (scale bars: 35 μm) (**B**). Histograms of the dry mass density values per quarter frame in one AGA tissue sample (**C**). Histogram of the dry mass density values per quarter frame in one SGA tissue sample (**D**). Gaussian fitting is used to separate the high and low myelin regions.

The cSLIM system was implemented using a commercial SLIM module (Cell Vista SLIM Pro, Phi Optics, Inc) attached to a commercial phase contrast microscope (Axio Observer Z1, Zeiss). The system was outfitted with a Zeiss EC Plan-Neofluar 40x/0.45NA objective and a Point Grey color camera. cSLIM is fully automated and operates as a whole slide scanner that provides simultaneously a standard brightfield image (e.g., of H&E, immunochemical stains, etc.) and a quantitative phase map[43].

The annular condenser ring is still employed and allows shifting of the unscattered light at the spatial light modulator (SLM) of the add-on module (Figure 1). For every phase shift, three intensity frames are acquired for each channel of the camera:

$$I(x, y; \varphi) = rR(x, r; \varphi) + gG(x, y; \varphi) + bB(x, y; \varphi), \quad (1)$$

where the weighing coefficients $r, g, b$ are related such that $r + g + b = 1$. For our measurements, we used $r = 0.1, g = 0.6, b = 03$, which was determined empirically by finding the equivalent central wavelength, $\lambda$, at which the phase modulation of the SLM occurs. The phase map is then computed from the four frames as previously described with SLIM[40].

The dry mass density $\rho$ of the tissue section is computed from cSLIM phase images using the following relationship:

$$\rho(x, y) = \frac{\lambda}{2\pi\eta} \varphi(x, y), \quad (2)$$

Where $\lambda$ is the center wavelength of the optical source, $\eta = 0.2\, ml/g$, corresponding to an average of reported values,

and $\eta = 0.2\,ml/g$, is the phase value of the tissue at different points. The resulting dry mass density has units of mass/area. Once the phase map of myelin alone was generated, the total dry mass density was calculated by integrating $\rho$ over all selected quarter frames.

We imaged the entire internal capsule from each slide. The data from each slide consisted of 625 cSLIM images, each the result of 12 intensity measurements (four phase shifted color images). To confirm that the entirety of the internal capsule was captured, a mosaic was constructed for each slide (Figures 2A-B). The regions adjacent to and surrounding the internal capsule were segmented out, because they contain primarily cell body content as indicated by the reduced LFB stain, prior to myelin content cSLIM quantification.

In order to separate the high and low myelinated regions, a histogram of dry mass for each quarter frame was evaluated, generating a bimodal pattern, which we described using gaussian fits, thereby selecting out the high myelin quarter frames from the rest of the sample (Figure 2C). After determining each quarter frame corresponding to the internal capsule, we applied a segmentation scheme to remove all the components of the tissue that do not correspond to myelin. The brightfield image for each frame was first used to generate a binary mask specific to myelin through a color threshold and a filtering function (Figure 3A-B). The SLIM counterpart image was then multiplied by the mask to produce a phase map specific to myelin (Figures 3C-D). All measurements were made while blind to slide-diet pairings.

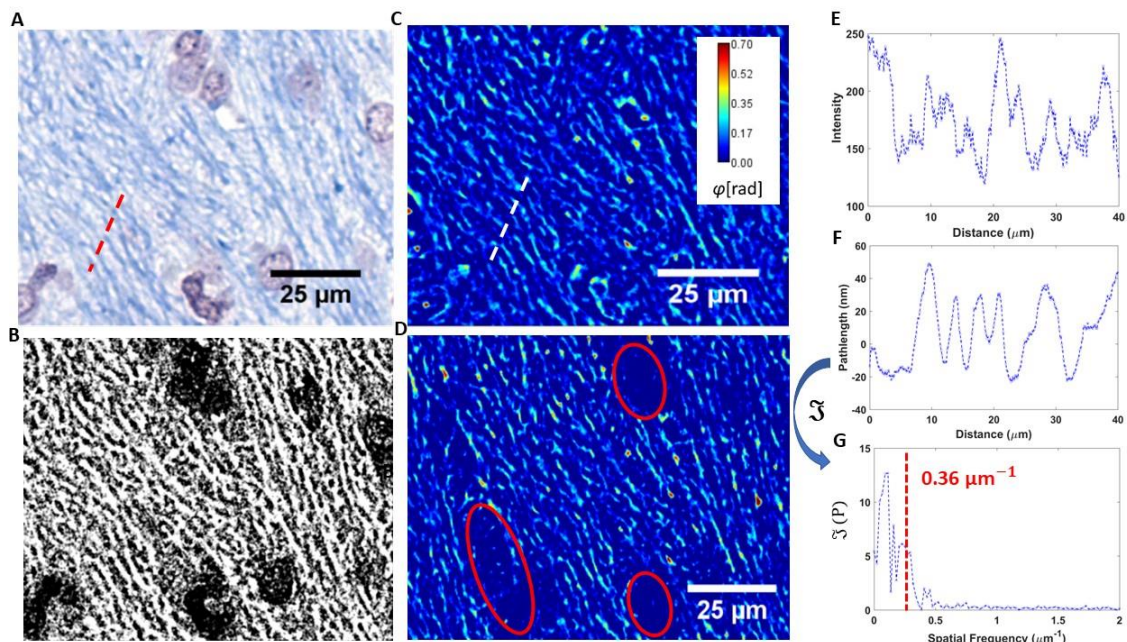

**Figure 3.** Myelin segmentation process. A brightfield image (**A**) is first used to create a binary mask specific to myelin (**B**), which is then multiplied by the corresponding SLIM image (**C**). The result is a phase map uniquely describing the myelin content of the image, with former cellular areas (encircled in red) deleted (**D**). Plot of the cross section in (A) (**E**). Plot of the cross section in (C) (**F**), and its Fourier transform (**G**), which has a mean frequency of 0.36 /µm, corresponding to an average axonal width of 2.7 µm.

Cross sections over myelinated axons were traced in both brightfield and corresponding SLIM images (Figure 3A, C). The resulting spectra of intensity and pathlength are shown in Figures 3E-F. Taking the Fourier transform and computing the mean spatial frequency yields a value of $0.36\ \mu m^{-1}$, translating to $2.7\ \mu m$ as the average external diameter of the myelin sheaths, which agrees well with previous findings[48].

Clearing brain tissue with the CLARITY method removes its lipids and therefore reduces drastically the myelin content. CLAIRTY was applied on coronal mouse brain (the piglet samples were permanently coverslipped) sections using 8% sodium dodecyl sulfate (SDS) according to a standard protocol[49]. Scans of 53 by 95 frames were captured and computationally stitched to form a mosaic image (Figures 4A-B). The anterior commissure and the corpus callosum were targeted for analysis (Figures 4A-B) because these regions are comprised primarily of myelinated axon tracts.

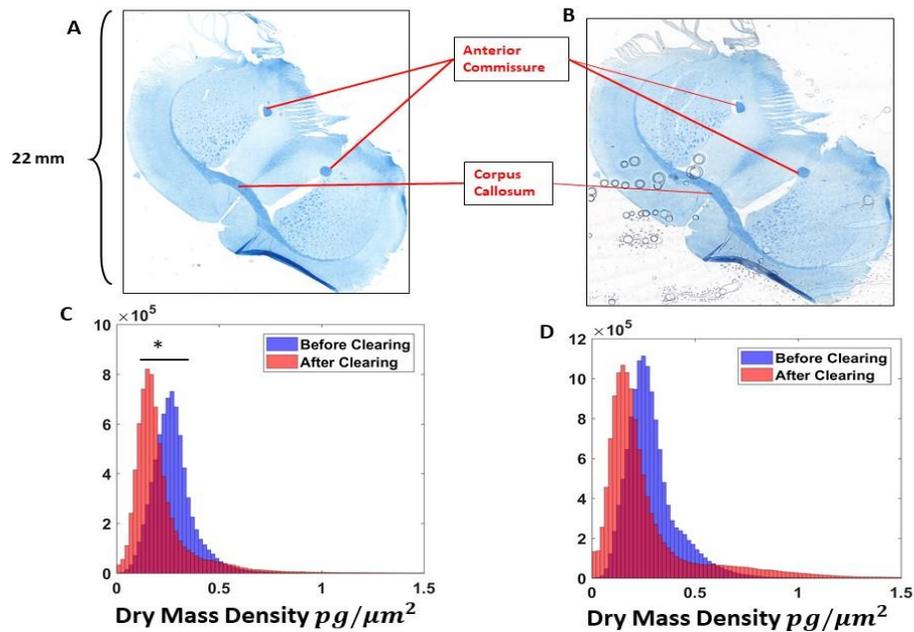

**Figure 4.** 53 x 93 frame stitches of a mouse brain tissue before (A) and after (B) the clearing procedure. Dry mass density histograms of the anterior commissure (C) portions and the corpus callosum (D) structure. (* p<0.05).

All processing was performed in MATLAB™. The statistical method employed to assess the significance of the results is the student's t-test. Significant results in the figures are indicated with an asterisk (*), which corresponds to a $p < 0.05$ Images of tissues and fiber tracking were prepared in ImageJ.

To test the ability of cSLIM to assess the dry mass content of myelin, we cleared mouse brain sections stained with LFB and imaged the specimens before and after clearing. The samples were scanned by cSLIM in 53 by 93 frames and the same segmentation scheme was used to process the images. Figure 4 shows a composite image of a coronal mouse brain section, stained with LFB, before and after the clearing process. The anterior commissure and corpus collosum, indicated in Figures 4 A-B, were chosen for comparative analysis. Contrasting the tissue before and after CLARITY revealed reductions in dry mass density measurements in both the anterior commissure and the corpus callosum, which was significant in the case of the anterior commissure, but not in the case of the corpus callosum (Figures 4C-D).

We evaluated samples from 4 treatment groups: AGA and SGA piglets on HF or CON diet. The regions analyzed were the internal capsule, which represented a myelin-rich and the corresponding surrounding parenchyma, which represented a low myelin brain region. Juxtaposing the brightfield images of SGA samples with those of AGA revealed that the four sets of samples differed in the depth of color and density of cells. The internal capsule contained darker, more fibrous structures, whereas the low myelin areas contained more numerous, larger cell bodies (Figure 5).

Significant differences were found between control and HF diet samples in all four tissue classifications, as shown in Figure 6. Myelin dry mass density in the internal capsule was markedly higher in AGA compared to SGA piglets, ($p < 0.01$). Similarly, higher myelin dry mass content was found in the surrounding parenchyma in AGA compared to SGA piglets (p<0.05). The mean values for dry mass surface density in the internal capsule was approximately $1\, pg/\mu m^2$ for the control AGA piglets, and approximately $0.8\, pg/\mu m^2$ for the control SGA piglets (Figures 6 A-B). Markedly lower average dry mass densities were found in the surrounding parenchyma for both AGA and SGA piglets (~ $0.5\, pg/\mu m^2$) (Figures 6C-D). These results are consistent with the previously mentioned work on these piglets using MRI.

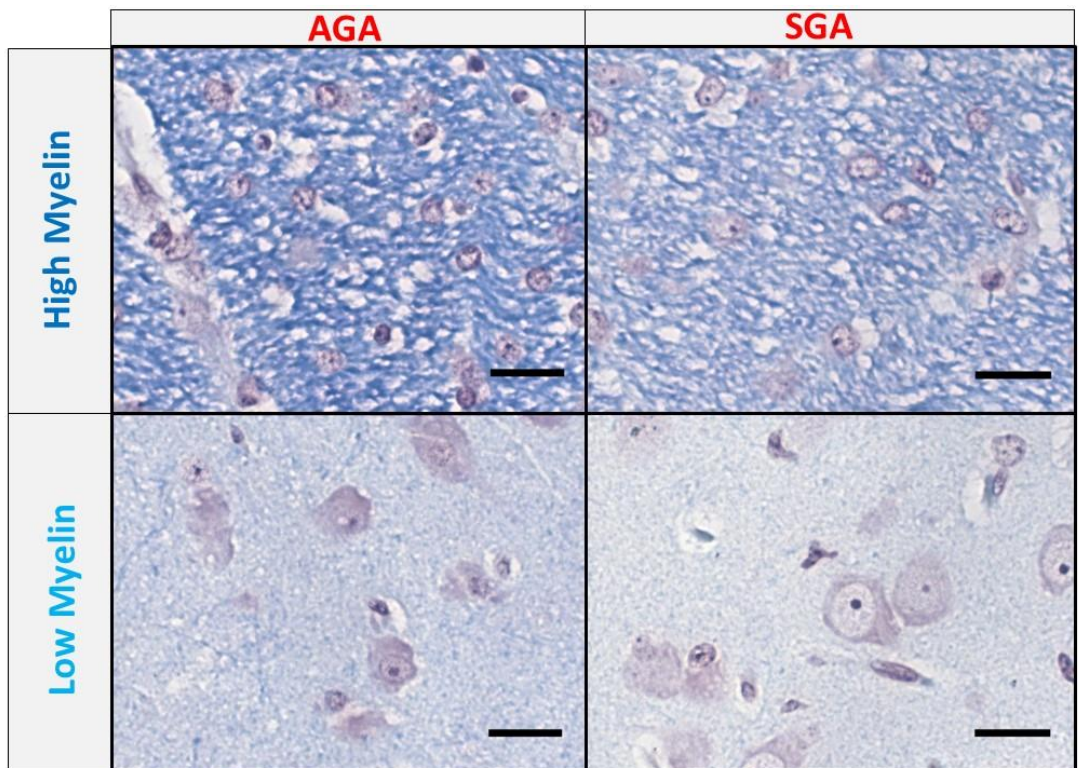

**Figure 5.** Single brightfield frames of high and low myelin areas in both appropriate for gestational age (AGA) and small for gestational age (SGA) samples.

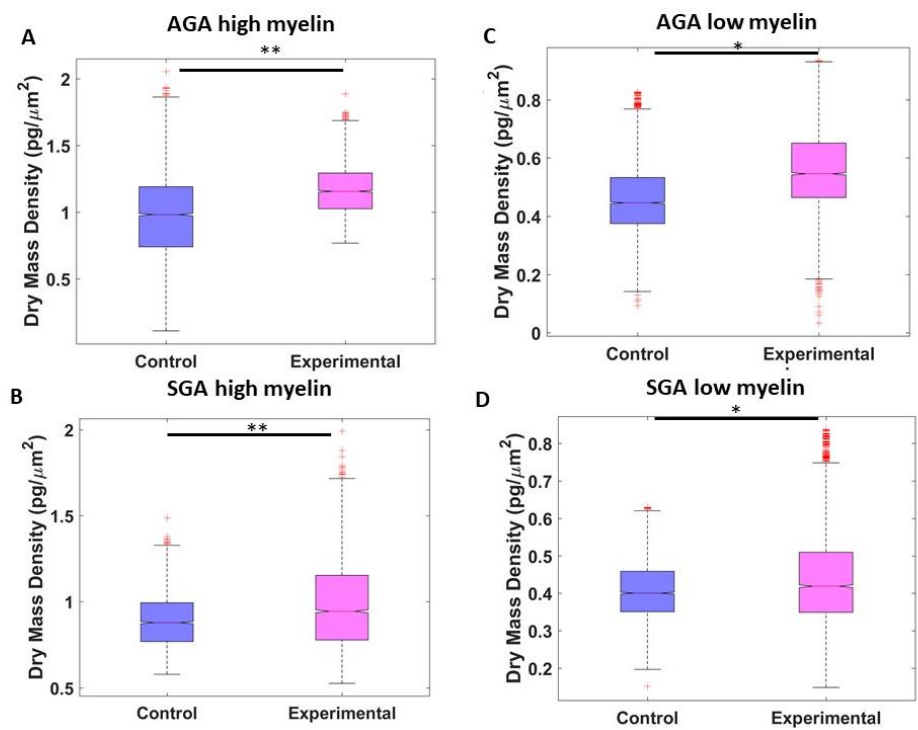

**Figure 6.** Difference in dry mass density per quarter frame of total AGA tissue with high myelin (A), SGA with high myelin (B), AGA with low myelin (C), and SGA with low myelin (D). (* $p<0.05$, ** $p<0.01$).

Our findings extend the field of analytical techniques that can quantify myelin in complex brain slices. Specifically, we present evidence that our method can detect differences in myelin dry mass using both the CLARITY method in mouse brain samples, and a nutritional intervention previously shown to impact myelination in the internal capsule of SGA and AGA piglet brains[43]. This is an important advance in the field given both the importance of myelination in brain development and function, and the difficulty of quantitatively measuring myelin in brain slices. To date, myelin measurement studies have been largely qualitative or have involved indirect quantifications and cumbersome, costly procedures. Novel quantitative approaches that are efficient and accurate are needed to enhance our understanding of deficient myelination and assess possible nutritional remedies.

We employed CLARITY to mouse brain samples to validate the ability of cSLIM to detect variations in myelination under controlled conditions that specifically removed lipids. Coupled with a segmentation based on the LFB stain, we measured myelin differences with great sensitivity. Our methodology successfully detected CLARITY-mediated decreases in myelin dry mass. These findings indicated that cSLIM is sensitive to myelin lipid loss following demyelination.

We then tested the ability of cSLIM to recapitulate previously reported[43] nutritionally and birth status driven differences in myelination in the internal capsule of piglets. These changes in myelin dry mass were substantially smaller in magnitude than the changes caused by CLARITY. Further, differences in myelination caused by birth status (SGA vs AGA) are likely to be qualitatively different from those caused by changes in dietary fat. We were able to reproduce the higher internal capsule myelin dry weight in AGA compared to SGA piglets using cSLIM in blinded fashion. We also reproduced the HF-mediated increase in internal capsule myelin dry weight in both SGA and AGA piglet brain samples. Additionally, we compared the dry myelin weight in the parenchyma surrounding the internal capsule to both the relevant internal capsule and across AGA and SGA and found substantially lower myelin density. Taken together, our results demonstrated that cSLIM was able to detect differences in myelin dry weight across a wide range of myelin concentrations in challenging biological tissues.

In summary, we have introduced a new technique for quantifying myelin content in brain tissue. Our approach involves the combination of the cSLIM imaging modality with a specificity mask provided by the simultaneous brightfield image of the stained tissue. To test our system, we imaged samples from piglets of different gestational ages fed diets with or without HF. We were able to recapitulate previous MRI findings using cSLIM in a blinded protocol. cSLIM can thus quantitate myelin content with high sensitivity and offers great promise for aiding the design of nutritional interventions for brain health.

We plan further studies to calibrate and fine-tune our assessment of myelin. For example, we plan to evaluate the absolute protein content of myelin at the single axon level by creating specificity masks using fluorescent tags for myelin proteins, such as the proteolipid protein. Co-culturing neurons and oligodendrocytes would allow an investigation into the dynamic formation of myelin around a single axon, in real-time. The development of technologies and imaging modalities for directly investigating myelination, deficiencies of myelin, and the processes of demyelination (disease) and remyelination (repair) in the context of dietary interventions is warranted given the substantial disability and high mortality rate associated with human myelin disorders, and particularly, those associated with IUGR and SGA.

## ACKNOWLEDGEMENTS

This work was supported by the National Institute of General Medical Sciences (NIGMS) grant GM129709, the National Science Foundation (NSF) grant CBET-0939511 STC, and the National Institutes of Health grant CA238191, as well as NRT-UtB 1735252.